\documentclass[10pt]{article}
\usepackage{multirow}
\usepackage[dvips]{graphicx}

\usepackage[psamsfonts]{amssymb}
\usepackage{amsxtra}
\usepackage{threeparttable}
\usepackage{amsmath}
\usepackage{amssymb}

\usepackage{graphicx}

\usepackage{cite}

\usepackage{color}

\usepackage{setspace}
\usepackage{float}
\usepackage{subfigure}

\usepackage[labelfont=bf,labelsep=period,justification=raggedright]{caption}

\makeatletter
\renewcommand{\@biblabel}[1]{\quad#1.}
\makeatother

\date{}

\begin{document}

\begin{flushleft}
{\Large
\textbf{Auditory Steady-State Response Stimuli based BCI Application - The Optimization of the Stimuli Types and Lengths}
}

Yoshihiro Matsumoto$^{1}$, 
Nozomu Nishikawa$^{1}$,
Takeshi Yamada$^{1}$,
Shoji Makino$^{1}$,
and
Tomasz M. Rutkowski$^{1,2,\ast}$
\\
\bf{1} Multimedia Laboratory, TARA Center, University of Tsukuba, Tsukuba, Japan\\
\bf{2} RIKEN Brain Science Institute, Wako-shi, Japan\\
$\ast$ E-mail: tomek@tara.tsukuba.ac.jp
\end{flushleft}

\section*{Abstract}

We propose a method for an improvement of auditory BCI (aBCI) paradigm  based on a combination of ASSR stimuli optimization by choosing the subjects' best responses to AM-, flutter-, AM/FM and click-envelope modulated sounds.
As the ASSR response features we propose pairwise phase--locking--values calculated from the EEG and next classified using binary classifier to detect attended and ignored stimuli.
We also report on a possibility to use the stimuli as short as half a second, which is a step forward in ASSR based aBCI. 
The presented results are helpful for optimization of the aBCI stimuli for each subject.

\section{Introduction}
Brain computer interface (BCI) is the technology which can allow for an operation of any computerized devices without any muscle activity~\cite{BCI-book}.
State of the art BCIs are implemented based on a classification of feature values extracted from brain activities, but their accuracies still not allow for broad applications. 
Therefore, there is a need for improvement of the classification accuracy for a better quality interfacing applications. 
Recent and hot research projects for the improvement are related to an auditory BCI (aBCI), which is a method applying auditory evoked responses (AER) generated by corresponding sound stimuli such as auditory steady-state response (ASSR)~\cite{ASSR}. 
Even though the contemporary studies report on relibable classification accuracy, there is still a plenty of
room for the improvement~\cite{Higashi}. 

\begin{figure}[H]
	\begin{center}
	\includegraphics[width=0.8\linewidth]{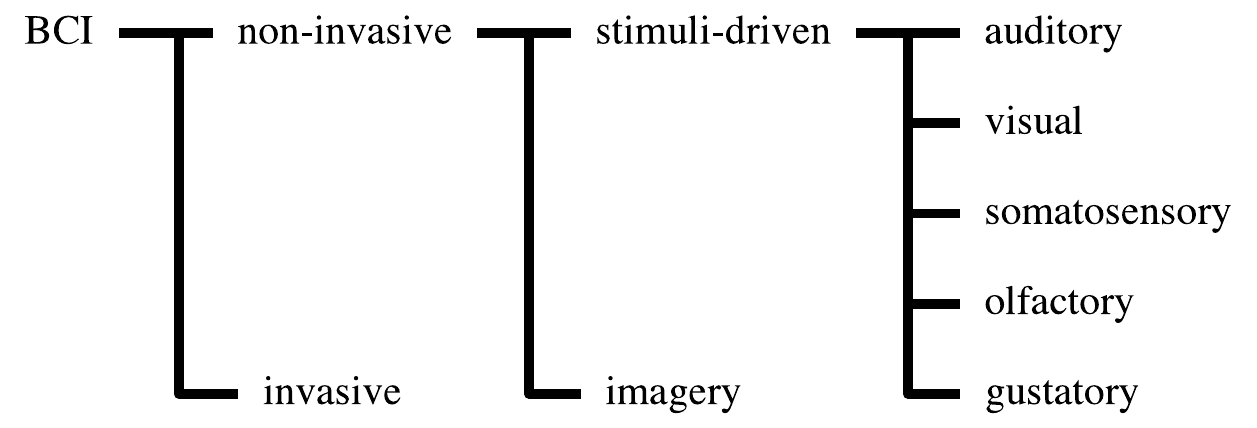}
	\end{center}
	\caption{Tree diagram of the BCI types. The BCI can be divided into invasive and non-invasive based on their implementation.
Moreover, the non-invasive BCI can be divided into imagery and stimuli-driven depending whether the sensory stimulus is used or not.
Following the general division of sensation into five categories, there are also five types of the stimuli-driven BCIs: visual; auditory; somatosensory; olfactory; and gustatory.}\label{fig:BCItree}
\end{figure}

There are several types in BCI shown as Figure~\ref{fig:BCItree}~\cite{BCI-book, yoshihiroB4thesis}.
The applications can be divided into two types: invasive- and non-invasive BCI.
Invasive BCI has usually a lower signal to noise ratio (SNR), but it requires a surgical operation of skull to embed the electrodes on/in cortical surface.
Contrarily, non-invasive BCI can be used by attaching electrodes on surface of human head, but the SNR is lower than in case of invasive technology.
Therefore, in case of the non-invasive BCI development of accurate applications is more challenging.
The next difficulty is related to simple interfacing paradigms design, of which imagery one based on user's intentional brain waves modulation caused by imagination/planning of certain actions (imagery movement, etc.) is the most popular one.
As a solution to this problem is a method which applies brain responses to artificial sensory stimuli for generating the commands called a stimuli-driven BCI (see Figure~\ref{fig:BCItree}). 
For example, an user chooses to concentrate on one stimulus from multiple presented sequentially, which evokes a specific event related potential (ERP) pattern called ``aha-response'' or P300, since it is a positive EEG deflection around $300$ms after the stimuli onset~\cite{book:eeg}. 
The BCI application classifies the responses and translates them into computerized commands. In case of aBCI
the ERPs evoked by the auditory stimuli are utilized.

The presented research focuses on the aBCI paradigm based on an auditory steady-state response (ASSR), which is usually used by  physicians as a method for a hearing level estimation~\cite{assr80hz}, and it is evoked by periodic modulated sound.
While a subject is listening the sound, the brain EEG response synchronizes with the amplitude modulating frequency.
Therefore, a spectral peak in EEG appears at the same frequency as the modulating one.
In addition, it was shown that while the subject concentrates listening to the sound stimulus, the response power rises~\cite{ASSR}.
The difference allows for a classification of the responses into \emph{targets} and \emph{non--targets}.
\emph{The target} refers to the chosen stimulus and \emph{the non--target} to the ignored one.

Applying this feature, Higashi et al.~\cite{Higashi} designed an aBCI, which was named an ASSR BCI.
That paper has preliminarily proven the efficiency of ASSR BCI, however, there is still necessity for further improvement before a successful application of the paradigms with patients in need.

In case of the classical ASSR based subjective hearing level estimation, there are studies showing that ASSR could be evoked by various modulation types, 
yet sinusoidal--amplitude--modulation (SAM) remains the most common choice in medical ASSR applications~\cite{ASSR}.
However, the hearing level estimation is often conducted while a subject is sleeping~\cite{ASSR,assr80hz}, thus there is no attentional related response modulation possible.
In other words, SAM sounds can cause large ASSR even when the subject is unconscious.
This feature is not desirable for the BCI application focusing responses to attentionally differentiated \emph{targets} and \emph{non--targets}.
Therefore, it is a need to further research on more efficient stimuli types for ASSR based BCI.

An additional problem that shall be resolved is related to the optimization of stimulation length. 
The stimuli should be long enough to evoke ASSR. 
However, if the longer stimuli are used, the inter-commands intervals will be also extended resulting in very low information--transfer--rates (ITR). 
Also the slower interfacing experience caused by longer stimuli interval downgrades the interfacing experience for the users.

Moreover, the longer stimuli interval would be an obstacle in development of an application based on multi-commands BCI.
Therefore, we aim to find the optimal stimulation length for the ASSR BCI.

In this paper we report on our research comparing the ASSR classification results for various sound stimuli types and lengths.

\section{Methods}

In this section, first we describe ASSR stimuli used for our BCI experiments. Next we discuss EEG signal processing and classification steps.

\subsection{The Conventional ASSR Stimuli}

\begin{figure}[H]
	\begin{center}
	\includegraphics[width=0.8\linewidth]{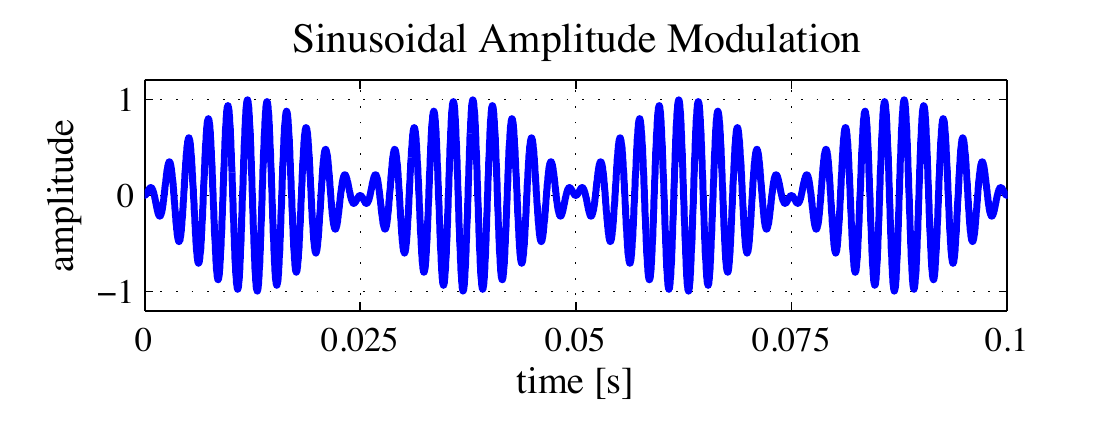}
	\end{center}
	\caption{A waveform of a $440$Hz tone amplitude modulated with a sinusoidal shape envelope with a frequency of $40$Hz.}\label{fig:SAM}
\end{figure}

The state of the art ASSR BCI, as reported in~\cite{Higashi}, could be built for a two commands interface based on SAM. SAM signal $s_1(t)$ is obtained as

\begin{equation}
	s_1(t) = \sin(2 \pi f_c t) \sin( \pi f_{m} t),
	\label{eq:SAM}
\end{equation}
where $t$ is a time-course, $f_c$ is a carrier tone frequency, and the $f_m$ is an envelope modulation frequency. An exemplary waveform is shown in Figure~\ref{fig:SAM}. 
In contemporary ASSR trials the stimulation length is usually $5$s or longer.

The conventional ASSR BCI application~\cite{Higashi} delivered the sound stimuli to the user from two different spatial directions one-by-one via headphones.
The directions corresponded to modulation frequencies (for example, $35$Hz stimulus from the left channel and $60$ Hz from the right one).
The user chose a command by concentrating on one of the directions, while ignoring the other one.
BCI application extracted the feature values from the monitored simultaneously EEG data. 
In a next step the features were classified leading to a resulting command. 
It was shown for the $10$ subjects that a classification accuracy was in a range $62\%$--$100\%$ for the binary case.
Those results are a proof of concept that the ASSR based BCI is a promising technology, yet the authors also pointed a necessity for further development.

\subsection{The Proposed Alternative ASSR Stimuli Types and Lengths}

We aim to improve the ASSR BCI by optimizing the stimuli types and their lengths. 
We test three new AM envelope types, which shall improve ASSR detection in EEG. 
The three new modulation envelopes are described in next sections and followed by a discussion of shorter stimuli lengths comparing to conventional ASSR applications.

\subsubsection{The novel envelope modulation types}
\begin{figure}[H]
	\begin{center}
	\includegraphics[width=0.8\linewidth]{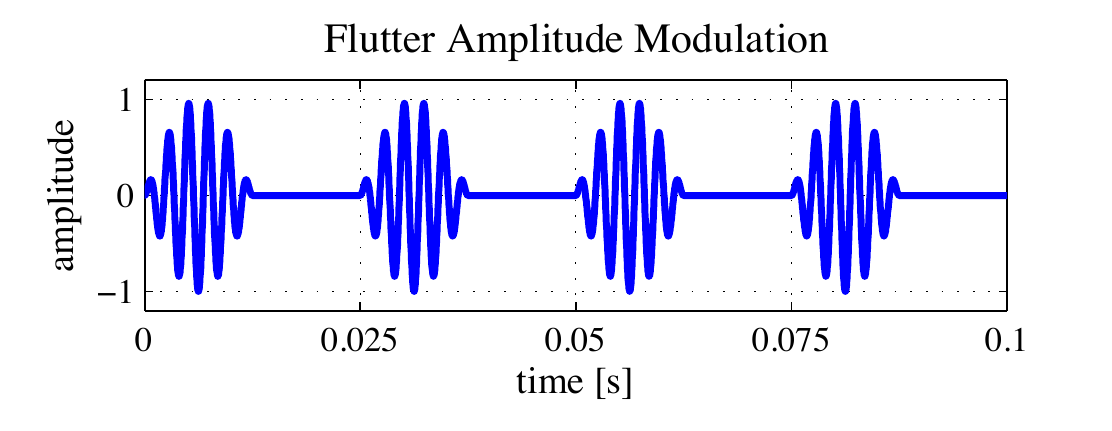}
	\end{center}
	\caption{A waveform of a $440$Hz tone amplitude modulated with a flutter shape envelope with a frequency of $40$Hz.}\label{fig:Flutter}
\end{figure}

We test the following three new modulations types: flutter AM (FAM), periodic clicks and combination of AM and FM (``AM/FM'') modulations.
The FAM stimulus $s_2(t)$ is obtained as follows,
\begin{eqnarray}
s_2(t)=
\left\{
\begin{array}{ll}
\sin(2 \pi f_c t) \sin(2 \pi f_{m} t) & (\sin(2 \pi f_{m} t) > 0)\\
0 & (otherwise)
\end{array}
\right.
\label{eq:Flutter}
\end{eqnarray}
where $f_c$ is the carrier frequency and $f_m$ is the modulating envelope one. 
The resulting exemplary waveform is depicted in  Figure~\ref{fig:Flutter}. 
Usually brain responses captured in EEG are easier to distinguish from more obvious changes like in case of pulses or transitions from silence to louder auditory sensation~\cite{book:eeg}. 
The FAM stimuli with silent intervals is based on the above observations. 
We expect the EEG features extracted from responses to the FAM stimulus to reflect the effect of surprising sound resulting with simpler classification of attended \emph{targets}. 

\begin{figure}[H]
	\begin{center}
	\includegraphics[width=0.8\linewidth]{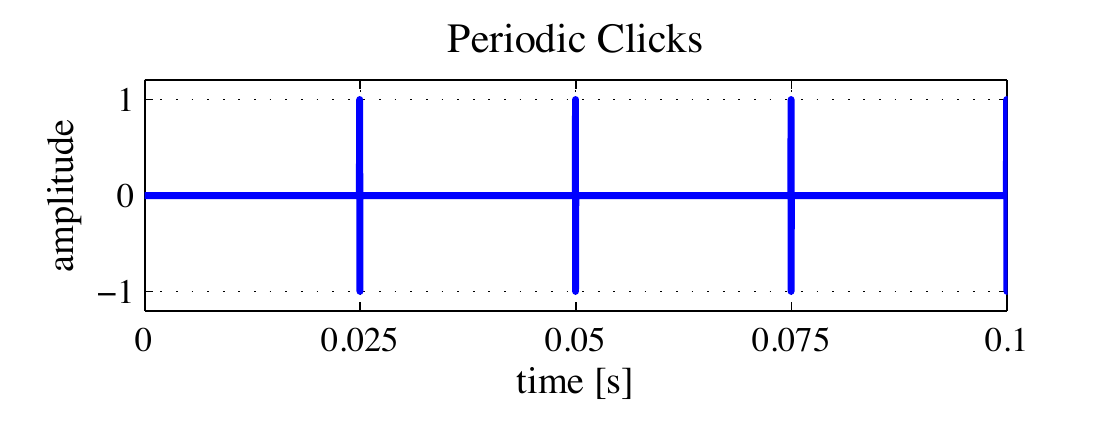}
	\end{center}
	\caption{A waveform of periodic clicks which have no carrier frequencies due to their very transient nature.}\label{fig:Clicks}
\end{figure}

A periodic clicks stimulus generates periodic click sounds with the same number of ticks as the frequency value.
The stimuli are generated from discrete train of neighboring $1$ and $-1$ values in each click interval as presented in Figure~\ref{fig:Clicks}.
In this paper, the frequency of clicks is defined as modulation frequency for easy comparison with the remaining ASSR stimuli types.
Click stimuli also have an effect as the surprising sound for the experimental subjects.
Moreover, the click stimulus is easier to attend or ignore by the subjects due to its very transient and broad frequency nature. 

\begin{figure}[H]
	\begin{center}
	\includegraphics[width=0.8\linewidth]{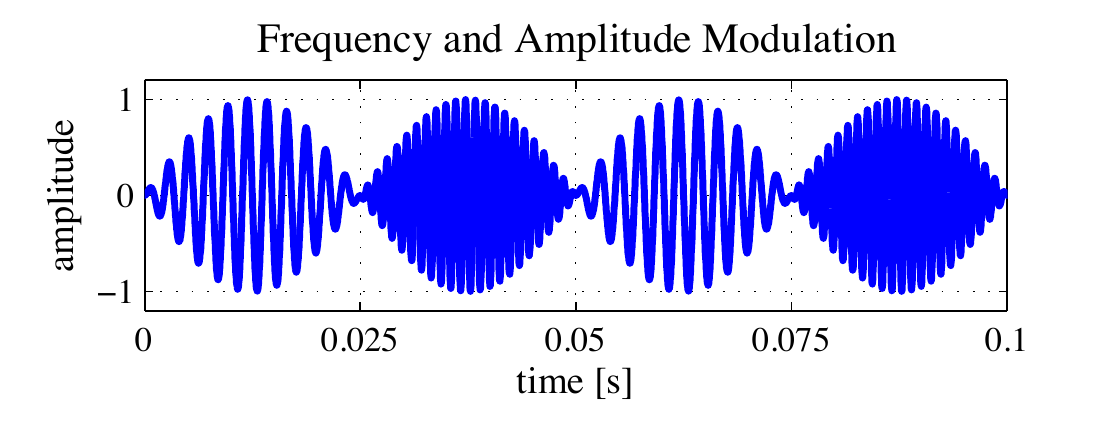}
	\end{center}
	\caption{A waveform as a combination of amplitude and frequency modulations of $440$Hz and $880$Hz tones.}\label{fig:FM}
\end{figure}

Finally we also propose the AM/FM based combination of signals to generate the ASSR envelope modulation based stimulus. 
This allows us also to check possibility to utilize frequency modulated tones. 
The AM/FM modulated ASSR stimulus $s_3(t)$ is generated as follows:
\begin{eqnarray}
s_3(t) = 
\left\{
\begin{array}{ll}
\sin(2 \pi f_{c1} t) \sin(\pi f_{m} t) & (\sin(\pi f_{m} t) > 0)\\
\sin(2 \pi f_{c2} t) \sin(\pi f_{m} t) & (otherwise)\\
\end{array}
\right.
\label{eq:AMFM}
\end{eqnarray}
where $f_{c1}$ and $f_{c2}$ are frequencies of the two carrier sounds,
as it can be seen in Figure~\ref{fig:FM}. 
The two different tones sound alternate synchronizing to create an AM envelope similar in shape as in case of the classical SAM stimulus in equation~(\ref{eq:SAM}).
AM/FM sound is more friendly for the subjects comparing to the simple SAM tone, since it has a more friendly alternating (musical) temporal structure. 
Therefore, we hypothesize, that application of this novel stimuli to the auditory BCI shall result in more comfortable interfacing experience.

\subsubsection{The various stimuli lengths}

The shorter stimuli for multi-stimuli paradigms, based alternating time sharing protocol, usual result with better ITR.
For example, time to present all commands in order to generate the final successful command becomes shorter.
That also results in lower burden or tiredness of the subjects.
On the other hand, if the stimuli would be too short, the machine learning algorithms would not be able to detect the ASSR.
We propose to evaluate the three stimuli lengths, which are still shorter comparing to the conventional lengths. 
The experiments conducted in this paper are with $0.5$s, $1$s and $3$s lengths.
In out preliminary experiment, reported in~\cite{yoshihiroB4thesis}, the $5$s long stimulation forced subjects to concentrate for a long time.
That resulted in increased tiredness and sleepiness causing a drop in interfacing efficiency.
Therefore, based on the above mentioned previous experience, in this paper we attempt only to use shorter stimuli intervals than 5s.
The comparison shall be helpful to optimize a tradeoff between the classification accuracy and the interfacing comfort for the subjects.

\subsection{EEG Experiment in Offline BCI Setting}
We conducted EEG experiments to record EEG for further offline analysis of the responses to various stimuli types and lengths.
The purpose is to find how much the changing experimental conditions affect the ASSR variability for  \emph{targets} versus \emph{non--targets} discrimination.

\subsubsection{EEG experimental settings}

The EEG experiments were conducted with five subjects who agreed to participate voluntarily. 
The experimental protocol was designed in accordance with our local institutions ethical committee and \emph{WMA Declaration of Helsinki} guidelines. 

In the experiments, the number of stimuli types was set to four as follows: SAM as in~(\ref{eq:SAM}), FAM as in~(\ref{eq:Flutter}), periodic clicks, and AM/FM as in~(\ref{eq:AMFM}).

The number of stimuli lengths was set to three: $0.5$s, $1$s and $3$s.

The stimuli were generated from three spatial directions which have corresponding modulation frequency: the $25$Hz modulated stimulus is always delivered only from the left channel; the $60$Hz one from the right; and the $40$Hz one from the both channels creating an illusion of a center location. 
The carrier tones for SAM and FAM stimuli were set to $440$ Hz. The AM/FM carrier tones were set to $440$ and $880$ Hz alternating tones.

\begin{figure}[H]
	\begin{center}
	\includegraphics[width=0.8\linewidth]{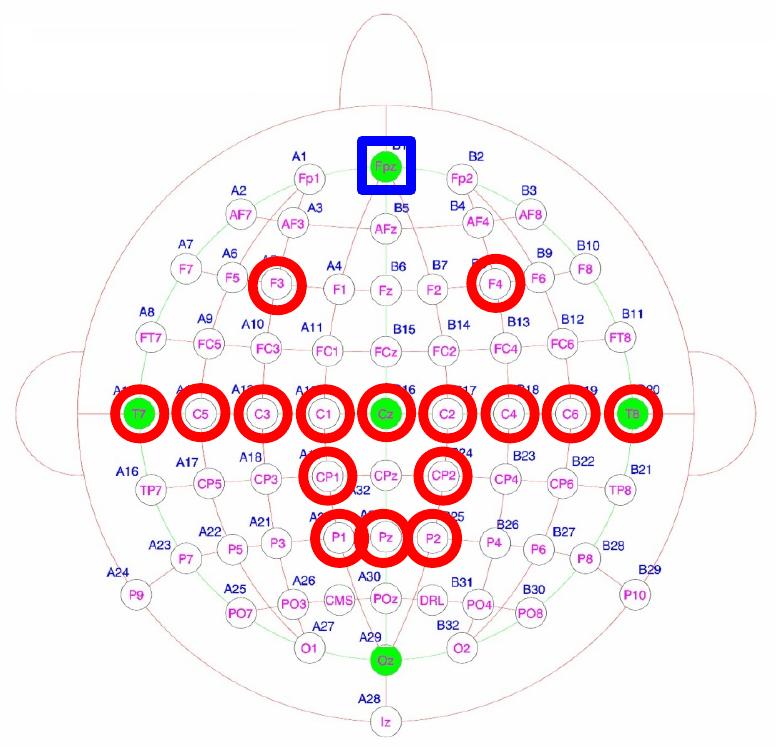}
	\end{center}
	\caption{The \emph{10/10 system} locations of EEG electrodes on a scalp. The graph is based on BIOSEMI's $64$-channels EEG cap layout.
The red circles represent recoding electrodes chosen for experiments reported in this paper. The blue square is a ground electrode. A reference was attached to the left earlobe.
}\label{fig:elec_pos}
\end{figure}

EEG was recorded using a $16$ channel \texttt{g.USBamp} amplifier by \texttt{g.tec}. The $16$ active wet electrodes were attached to the following scalp locations F$3$, F$4$, C$1$, C$2$, C$3$, C$4$, C$5$, C$6$. T$7$, T$8$, CP$1$, CP$2$, P$1$, P$2$, Pz, and Cz, as in \emph{10/10 system} layout~\cite{EEGpos} (see Figure~\ref{fig:elec_pos}). 
The reference was connected to the left earlobe. 
The ground was  connected to the FPz scalp location. 
The recorded EEG at each electrode was preprocessed by application \texttt{BCI 2000} as follows.
The sampling rate was set to $512$Hz and the  digital filters were set to pass EEG frequencies in a range of $5-100$Hz. 
A notch filter to remove electric power lines interference of $50$Hz was set in a band of $48-52$Hz according to the the East Japan power stations specifications.

\subsubsection{EEG experimental procedure}

\begin{figure}[H]
	\begin{center}
	\includegraphics[width=\linewidth]{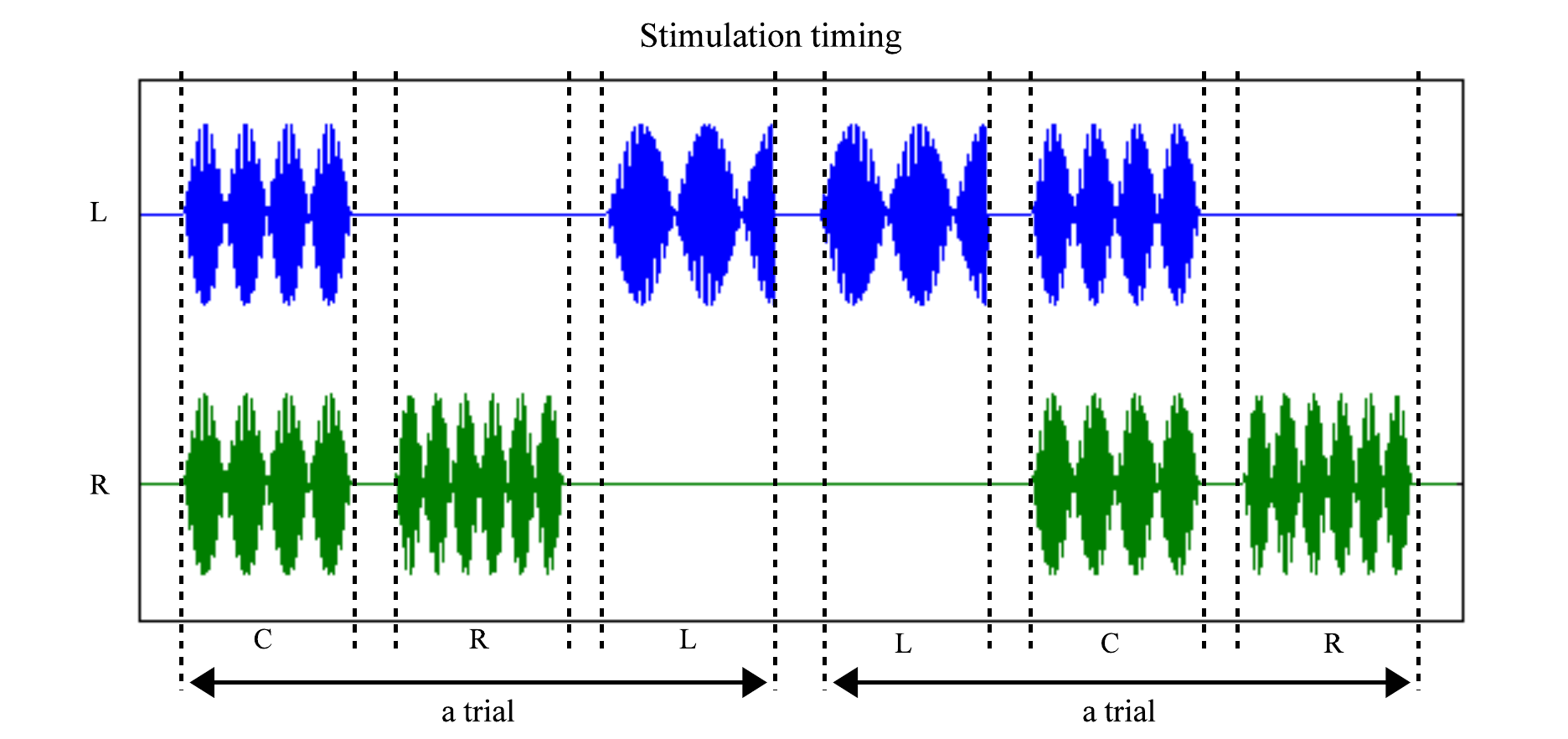}
	\end{center}
	\caption{The experimental protocol presenting the stimuli distribution in each trial. The horizontal axis represents the time-course.
The stimuli from a left audio channel are shown in blue and a right one in green. The stimulation spatial order in the figure in the first trial is center, next right, and finally left. In the second trial first left, next center, and finally right.}\label{fig:stim_t}
\end{figure}

The offline BCI experiment is conducted as follows.
Three frequencies stimuli are presented at random order from each of the three different directions via inner-headphones (see Figure \ref{fig:stim_t}). 
Presentation of the above three stimuli constitutes a single trial. 
We conduct experiments with each subject for $30$ trials.
During the first $10$ trials, a subject is instructed to concentrate only to the left direction with \emph{the target} stimuli and ignore the others as \emph{non--targets}.  
In the next $10$ trials - the center location is a target. Finally in the remaining $10$ trials - the right direction is a target.

The interval of breaks between each offset of a stimulus and onset of next one is set to $375$ms. 
There are also longer breaks between each of $10$ trials sessions. 
The above procedure is repeated for each of $12$ stimuli variation (three different lengths and four types).

\subsection{EEG Data Processing}

EEG data recorded on offline BCI mode during our experiments are next processed in order to extract features necessary for easy comparison of \emph{target} versus \emph{non--target}. 
The feature is necessary for a final classification leading to the BCI commands.

\subsubsection{ASSR features extraction from EEG}

As the feature value, we use a phase locking value (PLV)~\cite{PLV}. 
PLV outperforms the other tested methods of discrete Fourier transform (DFT) and coherence analysis.

PLV quantifies phase coupling between two signals which are recorded at different electrodes. 
When the two EEG signals are coupled, a hypothesis could be drawn that they originate from the same brain source or they are related to the same perceptual or cognitive process.
In case of the very noisy multichannel recordings such as EEG signals, a strong signal distributed among many electrodes is expected to result in multiple phase couplings. 
Unlike the amplitude based methods, such as coherence analysis, PLV can detect the signal which has a very tiny amplitude, like ASSR, buried in a very low SNR recording environment as in case of EEG captured from the human scalp.

PLV is obtained as follows.
At first, we apply a bandpass filtering to target only the interesting frequencies of the given signal $x_a(t)$, which is recorded at the $a^{th}$-electrode. 
The filter band is set from $(f_m-2)$Hz to $(f_m+2)$Hz. Next we construct an analytic signal for each electrode signal $x_a(t)$ as $\eta_a(t)$:

\begin{eqnarray}
\eta_a(t) = x_a(t) + i\tilde{x}_a(t),
\label{eq:eta}
\end{eqnarray}
where $i$ is an imaginary unit and $\tilde{x}_a(t)$ is the Hilbert transformed version of $x_a(t)$.
Next with an instantaneous phase $\theta_a(t)$ and a radius $R(t)$ the equation (\ref{eq:eta}) can be substituted by the following formulation:
\begin{eqnarray}
x_a(t) + i\tilde{x}_a(t) = R(t)\exp({i\theta_a(t)}).
\label{eq:Ax}
\end{eqnarray}
From to the equation (\ref{eq:Ax}), $\theta_a(t)$ which is an instantaneous phase of the original signal $x_a(t)$, is obtained as 
\begin{eqnarray}
\theta_a(t) = \arctan\frac{\tilde{x}_a(t)}{x_a(t)}.
\label{eq:arctan}
\end{eqnarray}
Finally, the difference between instantaneous phases of two signals $x_a(t)$ and $x_b(t)$ is defined as $\Delta\theta_{a, b}(t)$ and it is obtained as,
\begin{eqnarray}
\Delta\theta_{a, b}(t) = n\theta_a(t) - m\theta_b(t),
\label{eq:dtheta}
\end{eqnarray}
where $n$ and $m$ are ratio of synchronicity between $x_a(t)$ and $x_b(t)$.
In case of a single brain source detection from EEG, $n : m$ equals $1 : 1$, leading to $m = n = 1.$

Practically, the PLV is quantified by averaging of the unit vector $\exp({i\Delta\theta_{a, b}(t)})$ over time. Therefore, in case of discrete signals, PLV is obtained as
 \begin{eqnarray}
{\rm PLV}_{a, b} = \left|{1 \over L}\sum^{L}_{l=1}   \exp({i\Delta\theta_{a, b}(l)}) \right| ,
\label{eq:PLV}
\end{eqnarray}
where $L$ is the length of the analyzed signal window.
In our case $L$ equals to the stimulus length.
The PLV is normalized between zero and one.
The zero means that $x_a(t)$ and $x_b(t)$ are not coupled at all. 
The PLV equals to one means that the two analyzed EEG channels are perfectly coupled in the narrow analysis frequency band.

We calculate PLVs for each ASSR in EEG and for all pairwise combinations of the $16$ electrodes. As a result from the whole experimental session, we obtain $90$ ($3 \times 30$ trials) PLV sets from all stimuli types and lengths.
Each PLV set contains $120$ features $\left({{16}\choose{2}}~\mbox{combinations}\right)$, because ${\rm PLV}_{a, b} = {\rm PLV}_{b, a}$ and ${\rm PLV}_{a, a} = 1$.

Finally, we create a $120$-dimensional feature vector $\bold{f}$ for the classification of the each $k^{th}$ stimuli as follows:
 \begin{eqnarray}
\begin{array}{lllll}
\bold{f}_k = [ & PLV_{1,2}, &  PLV_{1,3}, & \ldots, & PLV_{1,16}, \\
& PLV_{2, 3}, & PLV_{2,4}, & \ldots, & PLV_{2,16}, \\ 
& \ldots, & PLV_{15, 16}].\\
\end{array}
\label{eq:fvec}
\end{eqnarray}

\subsubsection{Classification Method}

For the classification, we use a method of naive Bayes classifier (NBC)~\cite{book:pattRECO} applied in  \emph{leave--one--out} cross-validation mode using NaN-toolbox~\cite{NaNtoolbox}. 
The NBC method is rated as the best in comparison with linear discriminant and regression analysis based classification techniques~\cite{book:pattRECO}.

We contact a comparison among various stimuli types and lengths as in the following approach: 
\begin{itemize}
	\item First, we try to find out how much the \emph{target} versus \emph{non--target} detection is affected in various experimental conditions.
	\item Second, we compare whether we could discriminate targets based on different acoustic source locations.
\end{itemize}
 
The first test is to confirm ASSR difference between \emph{targets} and \emph{non--targets}.
This result is obtained from classification result in each stimulation with responses labeled as \emph{targets} and \emph{non--targets}, based on experimental instruction given to the subjects.
This test is based on a separation of the $90$ responses into three direction classes: left, right, and center.
We obtain $30$ samples for each direction class: $10$ samples are labeled as \emph{target}-direction and the remaining $20$ become the \emph{non--target}-directions.

The latter test is conducted to confirm the accuracy of ASSR BCI commands.
The responses from each single trial are labeled into three classes of: left, center, and right directions.
We combine the three feature vectors of each trial and redefine that as a single sample.
Therefore, there are $360$ features for each stimuli and $30$ samples for each type and length ($10$ samples labeled as left; $10$ as center; and $10$ as right).

In the both tests, the classifier for a single condition is independent from the others.

\section{Results}

For the comparison among different experiments, we use a value of classification accuracy (CA) resulting from the \emph{leave--one--out} cross-validation conducted in a single condition.
CA is defined as a rate of the number of the correctly classified ASSR responses to the all in a single experiment.

\subsection{Classification Results in \emph{target} versus \emph{non--target} Setting}

\begin{figure}[H]
	\begin{center}
	\includegraphics[width=0.8\linewidth]{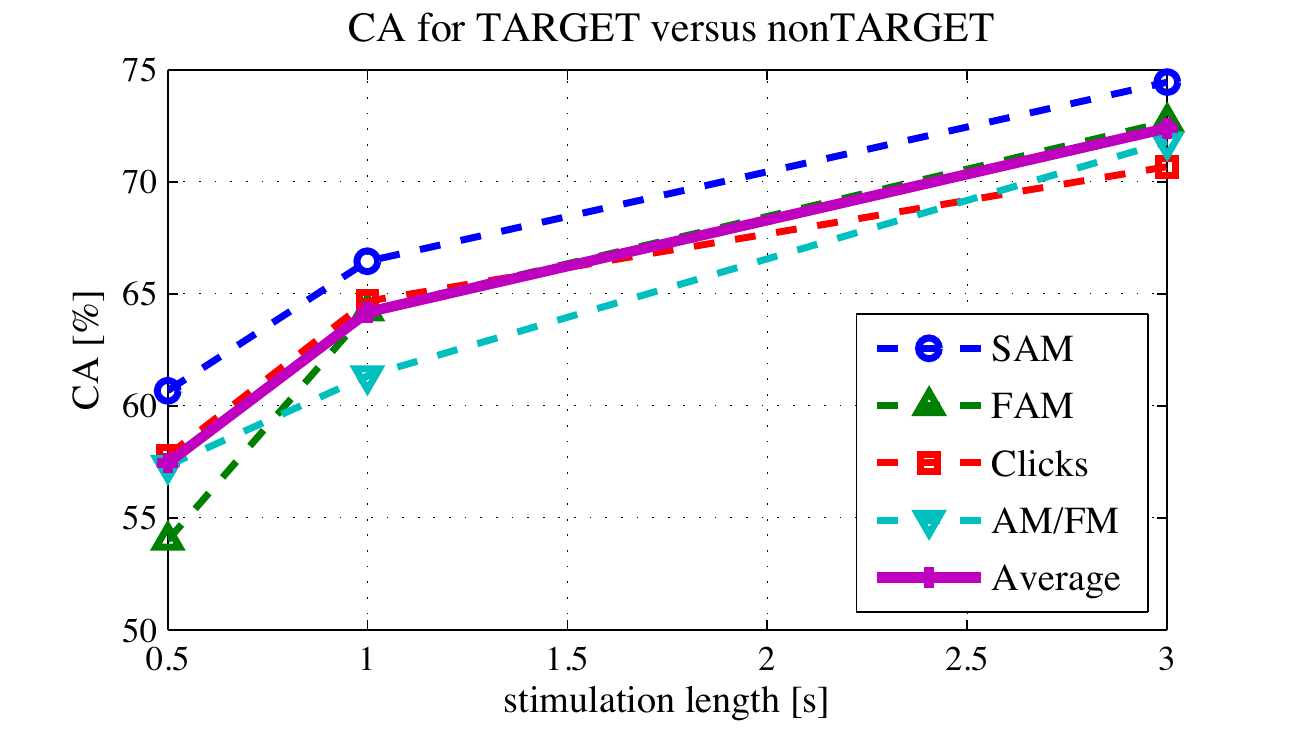}
	\end{center}
	\caption{Results of binary classification accuracies for \emph{target} versus \emph{non--target} setting for various stimuli types and an average of all subjects.}\label{fig:TNline}
\end{figure}

\begin{figure}[H]	
	\begin{center}
	\includegraphics[width=0.8\linewidth]{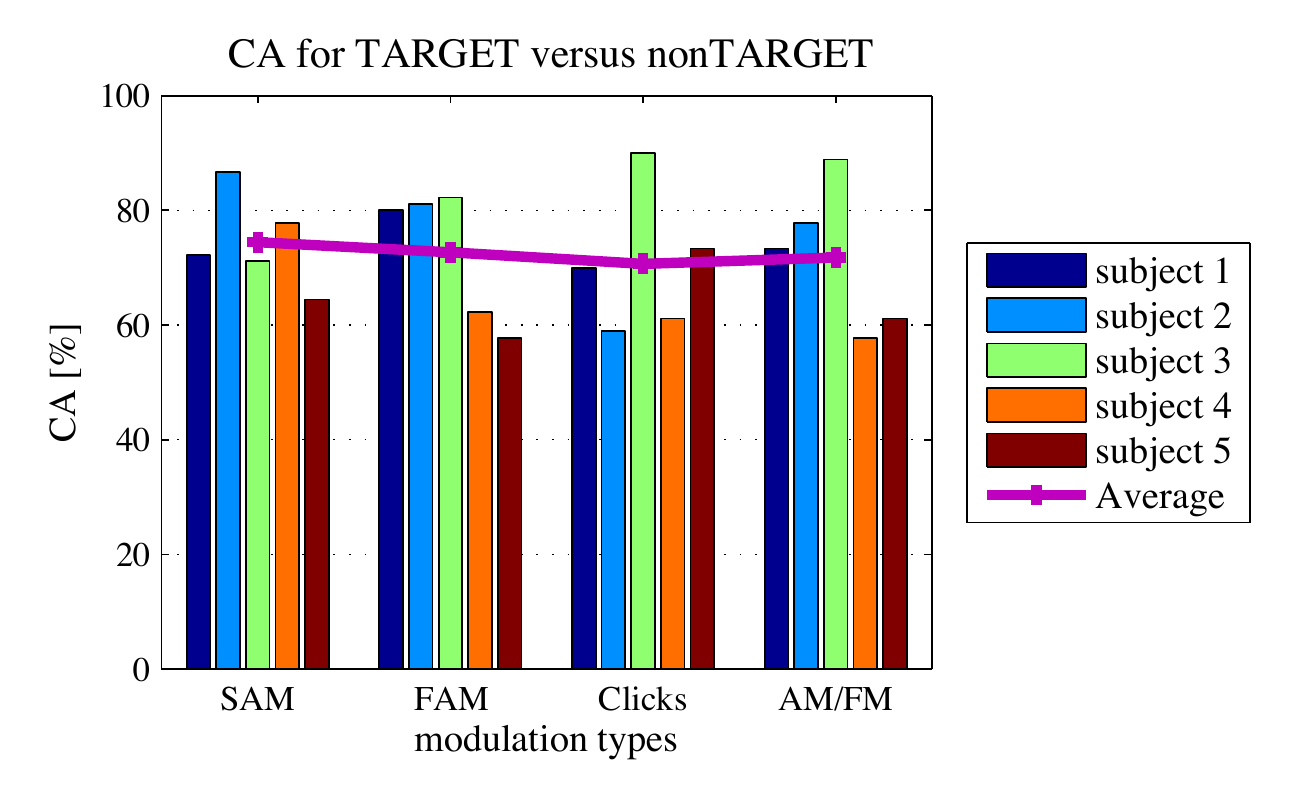}
	\end{center}
	\caption{A plot of classification accuracy based on modulation types for \emph{target} versus \emph{non--target}, for only $3$ long stimuli and for each subject separately.}\label{fig:TNbar}
\end{figure}

The results of classification for \emph{target} versus \emph{non--target} are shown in Figures~\ref{fig:TNline}~and~\ref{fig:TNbar}.
Figure~\ref{fig:TNline} contains a plot of CA averaged over all subjects for the comparison between stimulation lengths.
The detail values are shown in TABLE \ref{tab:TNline}.
According to the all lines in that figure, the longer stimulation, the better CA becomes, so shortening ASSR stimuli below $3$s is not advised from ASSR BCI. 
Additionally, the SAM stimuli seems to be the best in averaged responses.

Figure~\ref{fig:TNbar} presents a plot of CA presents only results from the best stimuli length of $3$s length for the comparison of various modulations.
The detail values are shown in TABLE \ref{tab:TNbar}.
The results show no significant differences among the tested ASSR generators of SAM, FAM, periodic clicks, and AM/FM.
This is a very encouraging news that ASSR could be evoked with various periodic sound stimuli.
That could enhance BCI interfacing experience by lowering possible weariness or habituation to a monotonic repetitive sounds. 

\subsection{Classification for Targets Direction}

\begin{figure}[H]	
	\begin{center}
	\includegraphics[width=0.8\linewidth]{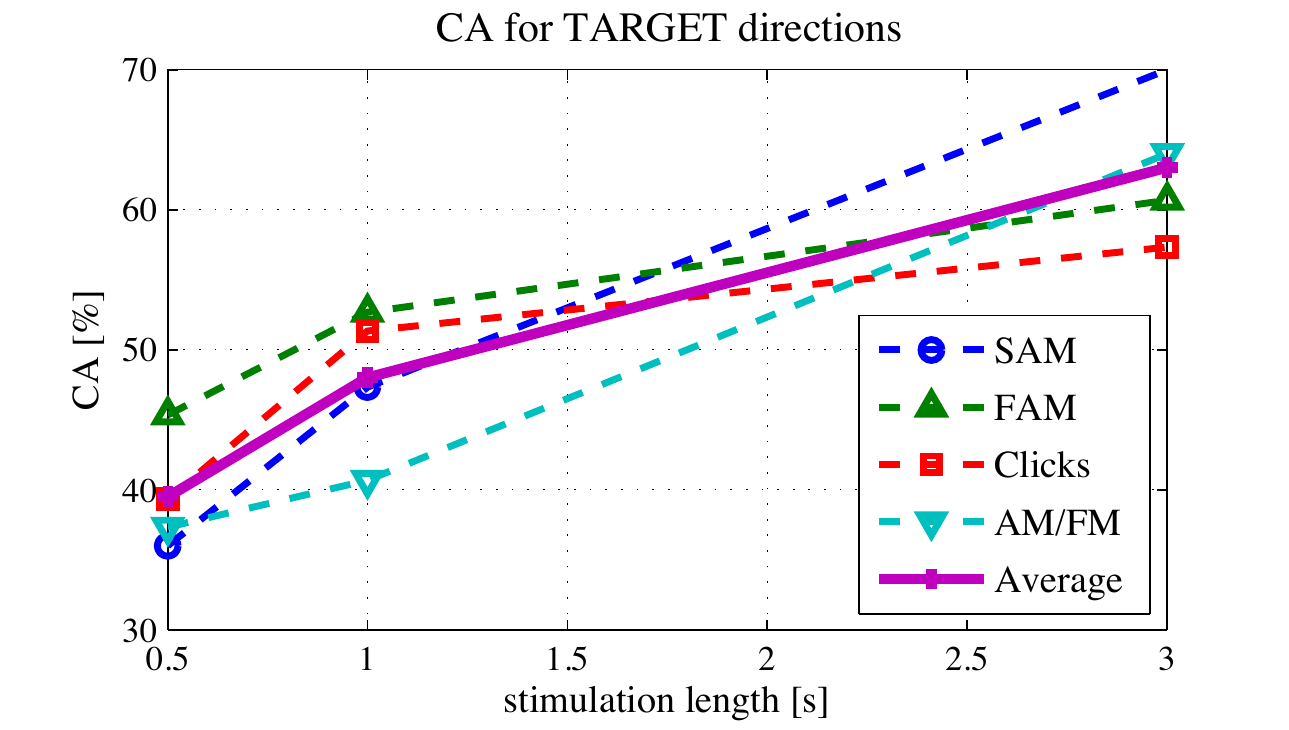}
	\end{center}
	\caption{A plot of classification accuracy based on stimulation length for \emph{target} directions with an average for all subjects.}\label{fig:CLline}
\end{figure}

\begin{figure}[H]	
	\begin{center}
	\includegraphics[width=0.8\linewidth]{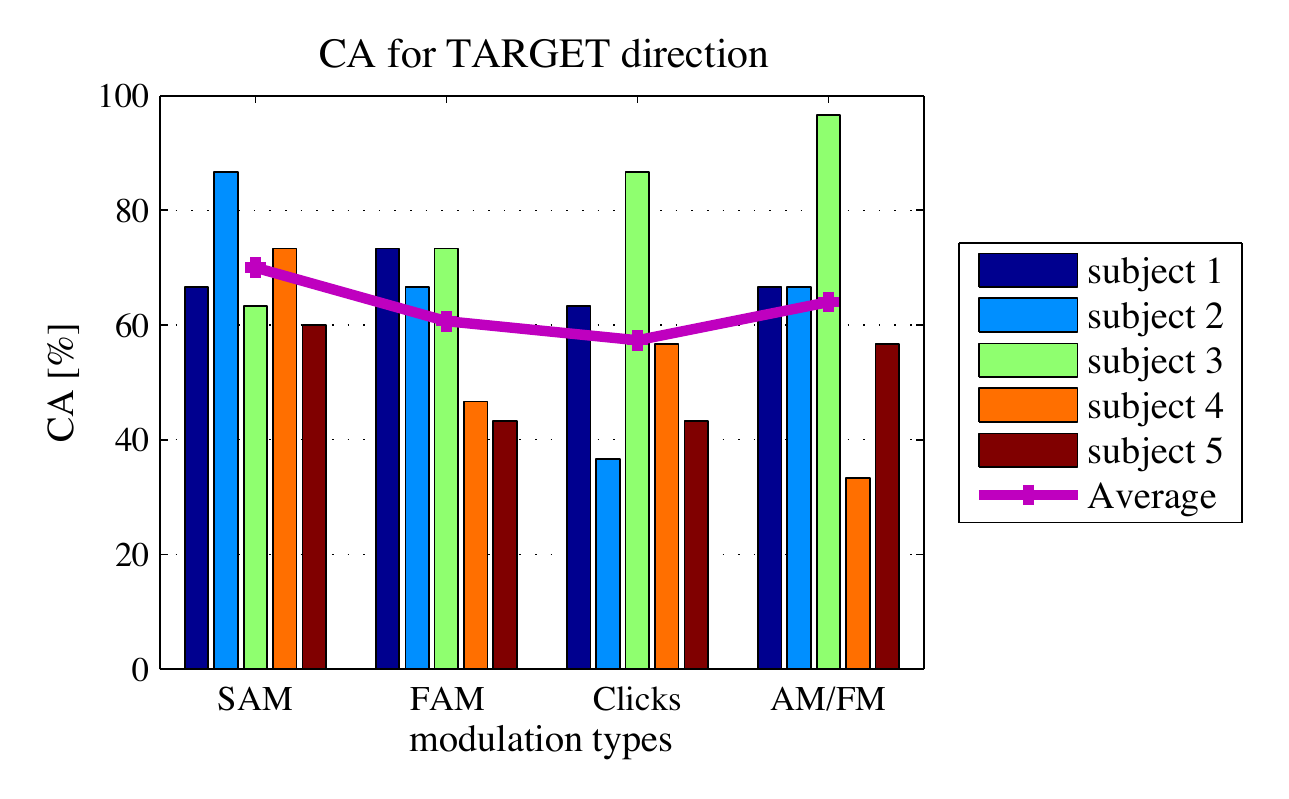}
	\end{center}
	\caption{A plot of classification accuracy based on modulation types for \emph{target} directions for only $3$ long stimuli and for each subject separately.}\label{fig:CLbar}
\end{figure}

The results of classification for \emph{target} direction within a single trial are shown in Figure~\ref{fig:CLline} for a comparison between stimulation lengths and in Figure~\ref{fig:CLbar} among various modulation types.
The detail values of Figure~\ref{fig:CLline} are shown in TABLE~\ref{tab:CLline} and those of Figure~\ref{fig:CLbar} are shown in TABLE~\ref{tab:CLbar}.
The same as in the above \emph{target} versus \emph{non--target} comparison results, Figure~\ref{fig:CLline} is created from the averaged results from all the subjects. 
Figure~\ref{fig:CLbar} is also generated from the $3$s long stimuli to compare stimuli types variability.
There is also no significant difference among the averaged results in Figure~\ref{fig:CLline}. 
On the other hand, the stimuli types cause differences in ASSR as shown by dotted lines in Figure~\ref{fig:CLline}.
According to Figure \ref{fig:CLbar}, the responses to SAM stimuli result with subject independent accuracies comparing to the other ASSR.
On the other hand, there are better stimulus types than SAM according to the accuracies for each subject.
When we average the accuracies over the best stimulus types for each subject, the result will be $78.00\%$.

\section{Conclusions} 

Based on the conducted EEG experiments in offline ASSR BCI mode, it has been shown that the stimulation length of $0.5$ second has been just above a chance level, as resulting from the NBC classification in  \emph{leave--one--out} cross-validation mode. 
Therefore, the $0.5$s stimulus is too short to cause ASSR possible to classify in a BCI setting.
On the other hand, the classification accuracy seems to increase monotonically together with the stimulation length.
The encouraging results for the $3$s long stimuli, where the accuracy seems to start to saturate already, are a step forward comparing with the state of the are ASSR BCI applications. 
More research is necessary in this ASSR stimuli range to determine with higher accuracy the optimal stimuli lengths.

A comparison of the various modulation types has been resulted with the outcome, that actually the conventional SAM based ASSR is the best for the three subjects out of the five tested. 
This result has shown that subject dependence is a very strong factor in aBCI design and the subjects shall be tested for their best and preferred responses.
According to presented results, while the average classification accuracy for SAM has been $70\%$, the best stimuli type for each subject has scored $78\%$ on average.
This performance is already adequate for practical BCI application, yet there is still plenty of room for possible improvement.

A single subject had better results for AM/FM and clicks based ASSR comparing to SAM. 
Those particular two stimuli types have different carrier frequencies comparing the the simple SAM stimuli. This results suggest that there is a possibility that the carrier sound could influence the results. 
The additional research topic shall be investigated to optimize the carrier sounds of the ASSR for each user.

Additionally, the modulation frequencies and stimuli directions has been unified for all the subjects in our experiments. 
The classification results could be improved by optimization of them for the subject. 
This will be also a topic of our future research.

On the other hand, optimizations for each user require long time for machine leaning before his/her first use of practical and successful online BCI.
Therefore, other approaches shall improve possibly user's convenience.

For example, the electrodes position could be further optimized, since in our experiments we also unified them.
In the presented study, the electrodes were distributed over the auditory cortices, temporal, parietal and frontal brain lobes. 
The PLV features has been calculated from the all electrodes combinations.
However, in case of  PLV, there is a danger that the very close positioned electrodes might results with unnecessary saturated values due to their close proximity.
Therefore, it is expected that more significant feature vector could be given by optimization of electrodes positioning.

In addition, we plan to extend or combine the ASSR with attentional alpha-wave modulation or synchronization with lower frequency steady-state responses, which already have shown in our former research~\cite{yoshihiroB4thesis} to have stronger modulatory impact. 
The other choice could be an inclusion of haptic/somatosensory or bone-conductance based stimulation based on tactile exciters attached to the fingers or facial areas.

The presented research is a step forward in development of novel BCI paradigms. 
We have confirmed and extended possibility to apply ASSR stimuli based on various stimuli for possible inclusion in practical online BCI applications.

\section*{Acknowledgments}

This research was supported in part by the Strategic Information and Communications R\&D Promotion Programme no. 121803027 of The Ministry of Internal Affairs and Communication in Japan, and by KAKENHI, the Japan Society for the Promotion of Science grant no. 12010738.

\newpage
\section*{Figure Legends}

\begin{description}
	\item[Figure~\ref{fig:BCItree}] Tree diagram of the BCI types. The BCI can be divided into invasive and non-invasive based on their implementation.
Moreover, the non-invasive BCI can be divided into imagery and stimuli-driven depending whether the sensory stimulus is used or not.
Following the general division of sensation into five categories, there are also five types of the stimuli-driven BCIs: visual; auditory; somatosensory; olfactory; and gustatory.
	\item[Figure~\ref{fig:SAM}] A waveform of a $440$Hz tone amplitude modulated with a sinusoidal shape envelope with a frequency of $40$Hz.
	\item[Figure~\ref{fig:Flutter}] A waveform of a $440$Hz tone amplitude modulated with a flutter shape envelope with a frequency of $40$Hz.
	\item[Figure~\ref{fig:Clicks}] A waveform of periodic clicks which have no carrier frequencies due to their very transient nature.
	\item[Figure~\ref{fig:FM}] A waveform as a combination of amplitude and frequency modulations of $440$Hz and $880$Hz tones.
	\item[Figure~\ref{fig:elec_pos}] The \emph{10/10 system} locations of EEG electrodes on a scalp. The graph is based on BIOSEMI's $64$-channels EEG cap layout.
The red circles represent recoding electrodes chosen for experiments reported in this paper. The blue square is a ground electrode. A reference was attached to the left earlobe.
	\item[Figure~\ref{fig:stim_t}] The experimental protocol presenting the stimuli distribution in each trial. The horizontal axis represents the time-course.
The stimuli from a left audio channel are shown in blue and a right one in green. The stimulation spatial order in the figure in the first trial is center, next right, and finally left. In the second trial first left, next center, and finally right.
	\item[Figure~\ref{fig:TNline}] Results of binary classification accuracies for \emph{target} versus \emph{non--target} setting for various stimuli types and an average of all subjects.
	\item[Figure~\ref{fig:TNbar}] A plot of classification accuracy based on modulation types for \emph{target} versus \emph{non--target}, for only $3$ long stimuli and for each subject separately.
	\item[Figure~\ref{fig:CLline}] A plot of classification accuracy based on stimulation length for \emph{target} directions with an average for all subjects.
	\item[Figure~\ref{fig:CLbar}] A plot of classification accuracy based on modulation types for \emph{target} directions for only $3$ long stimuli and for each subject separately.	
\end{description}

\newpage
\section*{Tables}

\begin{table}[H]
	\begin{center}
	\caption{Values of CA for \emph{target} verses \emph{non--target} based on stimulation length.}\label{tab:TNline}
	\begin{tabular}{| l || c | c | c |}
	\hline
	ASSR stimuli type & $0.5$s & $1$s & $3$s\\ \hline \hline
	SAM 	& $60.67\%$ & $66.44\%$ & $74.44\%$\\ \hline
	FAM  	& $54.00\%$ & $64.22\%$ & $72.67\%$\\ \hline
	Clicks      	& $57.78\%$ & $64.67\%$ & $70.67\%$\\ \hline
	AM/FM 	& $57.33\%$ & $61.33\%$ & $71.78\%$\\ \hline \hline
	Average of all stimuli  	& $57.44\%$ & $64.17\%$ & $72.39\%$\\
	\hline
	\end{tabular} 
	\end{center}
\end{table}

\begin{table}[H]
	\begin{center}
	\caption{The values of CA for \emph{target} verses \emph{non--target} based on modulation types.}\label{tab:TNbar}
	\begin{tabular}{| l | c | c | c | c |}
	\hline
	Subject	& SAM & Flutter & Clicks & FM\\ \hline \hline
	$\#1$ 	& $72.22\%$ & $80.00\%$ & $70.00\%$ & $73.33\%$\\ \hline
	$\#2$ 	& $86.67\%$ & $81.11\%$ & $58.89\%$ & $77.78\%$ \\ \hline
	$\#3$  & $71.11\%$ & $82.22\%$ & $90.00\%$ & $88.89\%$\\ \hline
	$\#4$ 	& $77.78\%$ & $62.22\%$ & $61.11\%$ & $57.78\%$\\ \hline
	$\#5$ 	& $64.44\%$ & $57.78\%$ & $73.33\%$ & $61.11\%$\\ \hline \hline
	Average of all	& $74.44\%$ & $72.67\%$ & $70.67\%$ & $71.78\%$\\
	\hline
	\end{tabular} 
	\end{center}
\end{table}

\begin{table}[H]
	\begin{center}
	\caption{Values of CA for \emph{target} directions based on stimulation length.}\label{tab:CLline}
	\begin{tabular}{| l | c | c | c |}
	\hline
	ASSR stimuli type & $0.5$s & $1$s & $3$s\\ \hline \hline
	SAM 	& $36.00\%$ & $47.33\%$ & $70.00\%$\\ \hline
	FAM & $45.33\%$ & $52.67\%$ & $60.67\%$\\ \hline
	Clicks & $39.33\%$ & $51.33\%$ & $57.33\%$\\ \hline
	AM/FM 	& $37.33\%$ & $40.67\%$ & $64.00\%$\\ \hline \hline
	Average of all stimuli & $39.50\%$ & $48.00\%$ & $63.00\%$\\
	\hline
	\end{tabular} 
	\end{center}
\end{table}

\begin{table}[H]
	\begin{center}
	\caption{The values of CA for \emph{target} direction based on modulation types.}\label{tab:CLbar}
	\begin{tabular}{| l | c | c | c | c |}
	\hline
	Subject & SAM & Flutter & Clicks & FM\\ \hline \hline
	$\#1$  	& $66.67\%$ & $73.33\%$ & $63.33\%$ & $66.67\%$\\ \hline
	$\#2$  	& $86.67\%$ & $66.67\%$ & $36.67\%$ & $66.67\%$ \\ \hline
	$\#3$   & $63.33\%$ & $73.33\%$ & $86.67\%$ & $96.67\%$\\ \hline
	$\#4$	& $73.33\%$ & $46.67\%$ & $56.67\%$ & $33.33\%$\\ \hline
	$\#5$ & $60.00\%$ & $43.33\%$ & $43.33\%$ & $56.67\%$\\ \hline \hline
	Average of all	& $70.00\%$ & $60.67\%$ & $57.33\%$ & $64.00\%$\\
	\hline
	\end{tabular} 
	\end{center}
\end{table}

\newpage
\section*{Figures}


\end{document}